\documentclass[final,5p,twocolumn]{elsarticle}

\usepackage{graphicx}
\usepackage{color}
\usepackage{amssymb}

\begin{document}
      
\title{WIMP-nucleon cross-section results\\ 
from the second science run of ZEPLIN-III}


\author[ITP]{D.Yu.~Akimov}
\author[ICL]{H.M.~Ara\'{u}jo\corref{cor1}}
\cortext[cor1]{Corresponding author}
\ead{h.araujo@imperial.ac.uk}
\author[EDI]{E.J.~Barnes}
\author[ITP]{V.A.~Belov}
\author[ICL]{A.~Bewick}
\author[ITP]{A.A.~Burenkov}
\author[LIP]{V.~Chepel}
\author[ICL]{A.~Currie}
\author[LIP]{L.~DeViveiros}
\author[RAL]{B.~Edwards}
\author[EDI]{C.~Ghag}
\author[EDI]{A.~Hollingsworth}
\author[ICL]{M.~Horn}
\author[ICL]{W.G.~Jones}
\author[RAL]{G.E.~Kalmus}
\author[ITP]{A.S.~Kobyakin}
\author[ITP]{A.G.~Kovalenko}
\author[ICL]{V.N.~Lebedenko}
\author[LIP,RAL]{A.~Lindote}
\author[LIP]{M.I.~Lopes}
\author[RAL]{R.~L\"{u}scher}
\author[RAL]{P.~Majewski}
\author[EDI]{A.St\,J.~Murphy}
\author[LIP,ICL]{F.~Neves}
\author[RAL]{S.M.~Paling}
\author[LIP]{J.~Pinto da Cunha}
\author[RAL]{R.~Preece}
\author[ICL]{J.J.~Quenby}
\author[EDI]{L.~Reichhart}
\author[EDI]{P.R.~Scovell}
\author[LIP]{C.~Silva}
\author[LIP]{V.N.~Solovov}
\author[RAL]{N.J.T.~Smith}
\author[ITP]{V.N.~Stekhanov}
\author[ICL]{T.J.~Sumner}
\author[ICL]{C.~Thorne}
\author[ICL]{R.J.~Walker}
\address[ITP]{Institute for Theoretical and Experimental Physics, 
Moscow, Russia}
\address[ICL]{High Energy Physics group, Blackett Laboratory, 
Imperial College London, UK}
\address[EDI]{School of Physics \& Astronomy, SUPA University of Edinburgh, UK}
\address[LIP]{LIP--Coimbra \& 
Department of Physics of the University of Coimbra, Portugal}
\address[RAL]{Particle Physics Department, 
STFC Rutherford Appleton Laboratory, Chilton, UK}

\date{\today}

\begin{abstract}
We report experimental upper limits on WIMP-nucleon elastic scattering
cross sections from the second science run of ZEPLIN-III at the Boulby
Underground Laboratory. A raw fiducial exposure of 1,344~kg$\cdot$days
was accrued over 319 days of continuous operation between June 2010
and May 2011. A total of eight events was observed in the signal
acceptance region in the nuclear recoil energy range 7--29~keV, which
is compatible with background expectations. This allows the exclusion
of the scalar cross-section above 4.8$\times$10$^{-8}$~pb near
50~GeV/c$^2$ WIMP mass with 90\% confidence. Combined with data from
the first run, this result improves to 3.9$\times$10$^{-8}$~pb. The
corresponding WIMP-neutron spin-dependent cross-section limit is
8.0$\times$10$^{-3}$~pb. The ZEPLIN programme reaches thus its
conclusion at Boulby, having deployed and exploited successfully three
liquid xenon experiments of increasing reach.
\end{abstract}

\begin{keyword}
ZEPLIN-III \sep dark matter \sep WIMPs \sep liquid xenon detectors
\end{keyword}

\maketitle

\section{Introduction}
\label{intro}

Direct, indirect and accelerator searches for neutralino dark matter
are now probing regions of parameter space favoured by minimal
supersymmetric (SUSY) extensions to the standard model, in particular
those constrained at the unification scale and by measurements of
cosmological cold dark matter abundance. SUSY is motivated by the need
to stabilise the weak scale, but it is remarkably persuasive that
$R$-parity conserving flavours of the theory lead to an excellent WIMP
dark matter candidate in the form of the lightest SUSY
particle. However, at a time when no evidence for SUSY has yet emerged
at the
LHC~\cite{atlas11,cms11,buchmueller11a,buchmueller11b,buchmueller11c},
it is worth noting that direct searches, such as the one reported
here, aim to detect any WIMP, not just neutralinos. Such experiments
exploit the possibility that WIMPs may scatter off ordinary baryonic
matter. The experimental challenge lies in conducting a rare event
search ($\lesssim$1~evt/kg/yr) whilst maintaining efficient detection
of very low energy signatures (few keV). Liquid xenon (LXe) is an
excellent target material for intermediate mass WIMPs due to its high
atomic mass and sensitivity in two response channels (scintillation
and ionisation). Significantly, these allow discrimination between
electron recoils resulting from radioactivity backgrounds and the
nuclear recoils expected from WIMP elastic scattering.

The ZEPLIN-III experiment operated at the Boulby laboratory (UK) under
a rock overburden of 2,850~m water equivalent. This two-phase xenon
emission detector measures both scintillation and ionisation responses
from particle interactions in its 12-kg LXe target. Approximately half
of this mass forms a `fiducial' region with well understood
performance and backgrounds.  The ionisation released at an
interaction site is drifted upward and emitted into a thin (few mm)
vapour phase above the liquid, where it is converted into an optical
signal via electroluminescence. This is achieved with a strong
electric field of 3--4~kV/cm in the liquid phase (approximately twice
as strong in the gas). An array of 31 photomultiplier tubes (PMTs) is
located within the cold liquid and views the 36.5-mm thick active
region above it. The array responds to the prompt scintillation and
the delayed electroluminescence signals (termed S1 and S2,
respectively). A detailed description of the detector design and
construction can be found in Refs.~\cite{akimov07,araujo06}. This time
projection chamber configuration allows very good position
reconstruction in three dimensions, as well as electron/nuclear recoil
discrimination which is critical for WIMP
searches~\cite{dolgoshein70,barabash89,bolozdynya95}.

The first science run (FSR) of the experiment in 2008 placed very
competitive upper limits on the WIMP-nucleon scattering cross sections
in several interaction models
\cite{lebedenko09a,lebedenko09b,akimov10a}. The FSR sensitivity was
limited by background originating from PMT $\gamma$-rays. In
particular, the most challenging event topology in ZEPLIN-III comes
from multiple-scintillation single-ionisation (MSSI) events, whereby a
single interaction vertex in the active region (producing both S1 and
S2) is accompanied by one or more scatters in a region yielding no
charge (S1 only). As the scintillation responses are effectively
time-coincident, S2/S1 ratios reconstructed for these $\gamma$-ray
events are essentially lower than typical for single-scatter electron
recoils, and they can leak down to the nuclear recoil acceptance
region. Nevertheless, we were able to achieve an average
electron/nuclear recoil discrimination power of 7,800:1 in the
2--16~keVee WIMP search region, which is the best reported for a LXe
detector (hereafter, `keVee' represents the electron-equivalent energy
as calibrated by 122~keV $^{57}$Co $\gamma$-rays and `keVr' denotes
nuclear recoil energy).

Two upgrades of the experiment had been planned from its
inception. The first was the replacement of the PMT array, which
dominated the $\gamma$-ray and neutron background budgets in the FSR
by a large factor. A new PMT model (ETEL D766Q~\cite{etel}) was
developed in collaboration with the manufacturers, which delivered a
40-fold improvement in $\gamma$-ray activity per unit. This allowed an
18-fold reduction in overall electron recoil background at low
energies relative to the FSR. In Ref.~\cite{araujo11} we analysed the
radioactivity backgrounds affecting the experiment in the second run
and showed that they were predicted with good
precision. Unfortunately, the optical and electrical performances of
the new PMTs were substantially poorer than those of the previous
tubes; worse still, the dispersion of gains and quantum efficiencies
(QE) posed very considerable problems to data analysis. Obtaining a
set of working PMTs and coping with this variability across the array
became the main challenge of the second run.

The second major upgrade was the addition of an anti-coincidence veto
system, which was retrofitted around the ZEPLIN-III
target~\cite{akimov10b,ghag11}. This 52-module plastic scintillator
detector envelops a Gd-loaded polypropylene shield ($\approx$3$\pi$
coverage) which provides moderation and radiative capture of internal
neutrons; the mean capture time is 10.7$\pm$0.5~$\mu$s. The assembly
fits inside the lead shield used in the FSR. The veto provides 60\%
tagging efficiency for internal neutrons, mostly derived from a
0.2--70~$\mu$s delayed coincidence window from the ZEPLIN-III trigger
point (58\%); elastic recoils in the plastic scintillator within a
narrow prompt window ($\pm$0.2~$\mu$s) make up the remaining neutron
efficiency. The tagging efficiency for $\gamma$-rays in the prompt
window was 28\%. Adding to its background rejection capability, this
tonne-scale detector provides a useful source of diagnostic for the
radiation environment around the instrument.

Minor upgrades were also implemented to aid with calibration of
ZEPLIN-III. The radioactive source delivery system was fully
automated. A copper structure (`phantom' grid) was installed above the
anode mirror; this cast a shadow from $^{57}$Co $\gamma$-rays onto the
LXe surface, thus providing calibration of the position reconstruction
algorithm. A new fibre-coupled LED light gun helped with calibration
of the PMT single photoelectron responses. All of the above upgrades
were manufactured from low background components.

\section{Second science run}
\label{SSR}

In the second science run (SSR), WIMP-search data were acquired over
319 days between 24$^{th}$ Jun 2010 and 7$^{th}$ May 2011, giving a
fiducial exposure of 1,344~kg$\cdot$days. A 20\% reduction from the
FSR fiducial mass to 5.1 kg was motivated by the poor performance of
peripheral PMTs. Even so, the SSR accumulated 3 times more exposure,
and achieved the longest continuous WIMP run of a xenon detector to
date. A daily operational duty cycle of 96\% was achieved
consistently, with 1~hr per day reserved for $^{57}$Co calibration and
cryogen re-filling; these tasks were automated and controlled
remotely. On a weekly basis, the system levelling was adjusted (to
compensate for local geological movement) and the target and veto PMTs
were calibrated with their respective fibre-coupled LED systems. The
gas phase was $\sim$3.5~mm thick and kept at 1.6~bar, with 13~mbar rms
variability within the dataset. The electric field in the liquid was
3.4~kV/cm. A free electron lifetime of 14~$\mu$s was achieved by prior
purification with a hot getter (comparable to the 13-$\mu$s drift time
for the deepest fiducial interactions). As observed in the FSR, the
lifetime increased steadily without external purification during the
run, and eventually reached $\sim$45~$\mu$s.

Data acquisition procedures were described with the FSR results
\cite{lebedenko09a} and we provide only a brief summary here. The
ZEPLIN-III trigger is derived from the 31-PMT sum signal; the SSR
hardware threshold corresponded to the electroluminescence of
$\approx$5 electrons; this translates to $\approx$10 ionisation
electrons in the liquid (average over fiducial interactions in the
dataset) when the finite electron lifetime and 66\% emission
probability at the surface are accounted for. Single emitted electrons
generate a mean response of 11.8$\pm$0.4 photoelectrons (phe); a study
of the single electron signature can be found in
Ref.~\cite{santos11}. This S2-derived signal also triggers the veto
data acquisition. The two data streams are recorded and reduced
separately, and synchronised offline. The ZEPLIN-III waveforms are
digitised at 500~MS/s for $\pm$18~$\mu$s around the trigger point. Key
detector and environmental data from the `slow control' acquisition
system were embedded into the main data.

\subsection{Data analysis}

Pulse-finding and parametrisation of the waveforms were carried out by
ZE3RA~\cite{neves11}, our event reduction software. The reduced data
were then searched for single scatter events (single S1 and S2 pulses)
which were retained for further analysis. Several corrections were
applied, mainly to the S2 response, based either on the slow-control
information embedded with each event, or on historical trends derived
from the daily calibration. The mean electron lifetime correction was
37\%, averaged over the fiducial dataset; also corrected were the
electronics gain drift (4.7\% rms), detector tilt (1.9\%) and pressure
variations (1.1\%).

A vertex reconstruction algorithm estimated the energy ($E$) and
position ($x$,$y$) of the interactions, with a maximum likelihood fit
to the S1 response and a least squares fit to the S2
channel~\cite{solovov11}. The algorithm fits to 31 empirical response
functions simultaneously. These model the response of a PMT as a
function of distance to its axis, and are derived from calibration
data in an iterative procedure which also `flat-fields' the array. The
spatial resolution thus achieved was 13~mm in S1 and 1.6~mm in S2
(FWHM) for the inner 100~mm radius. The ($E$,$x$,$y$) parameter space
was navigated with a {\em simplex} method to obtain the best fit point
for each event; spatial maps of likelihood and $\chi^2$ could also be
produced to help identify multiple vertices.

Finally, fiducialisation and quality cuts were applied to the data. A
central region of 140~mm radius containing 5.1~kg of LXe was
retained. The quality cuts eliminate likely MSSI events and outliers
in several parameters (e.g.~goodness-of-fit of S1 and S2
reconstructions). An S1 pulse timing cut was also used to exploit
modest pulse shape discrimination at the higher energies
~\cite{akimov02}. The inefficiencies incurred from these cuts are
described in the next section. Unsurprisingly, they are more severe
than had been required for FSR data.

A blind analysis was conducted during the initial optimisation of the
quality cuts and definition of the WIMP acceptance region: low energy
events in and around the signal region were kept hidden from
visualisation and reduction using a `blindness manager' implemented in
ZE3RA. Vetoed events (mostly $\gamma$-rays) were excluded from this
list, thus providing a (practically) signal-free population of
background events. This blind analysis was not pursued to its ultimate
conclusion but the acceptance region (which had been defined blindly)
was retained in order to avoid major bias in estimating a potential
signal.

\subsection{Calibration and efficiencies}

\begin{figure}[ht]
  \begin{center}
    \includegraphics[width=8.5cm,clip=on]{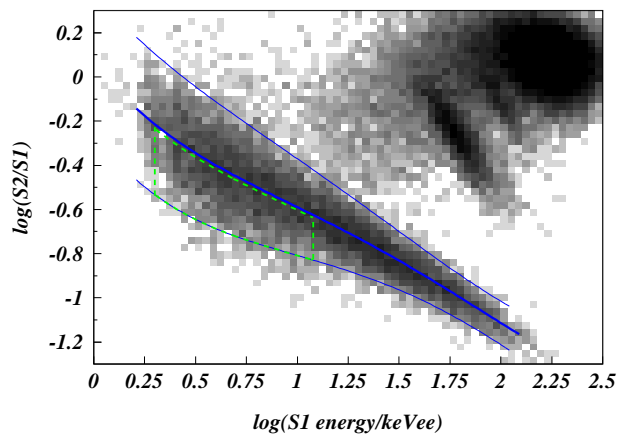}
  \end{center}
  \caption{\label{ambe} Discrimination parameter as a function of
  S1-derived energy for Am-Be neutron calibration. The recoil
  population median and $\pm$2-$\sigma$ lines are shown (in blue), along
  with the 2--12~keVee search box (dashed green).}
\end{figure}

Daily calibration with $^{57}$Co $\gamma$-rays (mostly 122~keV)
defined the S1 and S2 energy scales and monitored their stability
throughout the run; it also provided a regular measurement of the
detector tilt (through the spatial dependence of the S2 pulse width)
and of the free electron lifetime (by S2/S1 ratio with interaction
depth). The latter parameter was confirmed independently with a new
method based on single electron counting operating on the actual
science data waveforms~\cite{santos11}.

In the SSR the light yield of the chamber (fiducial average) was
1.3~phe/keVee, down from 1.8~phe/keVee in the FSR. The energy
resolution was similarly affected; a linear combination of S1 and S2
responses which exploits their microscopic anti-correlation yielded
16.4\% (FWHM) for the new array, against a very impressive 8.4\%
achieved for FSR data using the same analysis codes. In thin LXe
targets like ZEPLIN-III individual PMTs collect a large fraction of
the scintillation light of an event; the uniformity of response in the
array becomes therefore more critical than in high-reflectance
chambers with deeper geometries. In spite of this, a spatial
resolution of 1.6~mm (FWHM) in the horizontal plane was achieved in
the SSR, measured from the shadow pattern cast by the new phantom grid
for $^{57}$Co $\gamma$-rays~\cite{solovov11}.

The electron-recoil population was calibrated with a 4.6~kBq
$^{137}$Cs source located above the instrument, producing a rate of
150~c/s in the detector. The number of low-energy events thus obtained
was similar to that of background electron recoils in the search data.

Three neutron calibrations took place during SSR data-taking,
totalling 10~hours of exposure to an Am-Be ($\alpha,n$) source
emitting 1,321$\pm$14~n/s~\cite{NPL}; absolute differential rates of
nuclear recoils from elastic scatters agreed within statistical errors
for these datasets. The delayed-coincidence veto efficiency for such
events was 58\% independently of recoil energy~\cite{ghag11} (the
delayed detection is provided by $^{158}$Gd $\gamma$-rays rather than
by the neutrons directly). One such run is depicted in
Figure~\ref{ambe}. The recoil band populated by neutron elastic
scattering was parametrised by Gaussian fitting to the discrimination
parameter $\log_{10}$(S2/S1) in 1~keVee (S1) bins. Energy-dependent
mean ($\mu$) and width ($\sigma$) parameters were used to define a
signal acceptance region in the range 2--12~keVee and including
approximately the lower half of signal acceptance as indicated in the
same figure: from 2.3\% ($\mu-2\sigma$) to 45\%
($\mu-0.126\sigma$). This region was defined after a gradual
unblinding of surrounding data in the WIMP search dataset, with outer
regions progressively discarded and then opened for analysis as
`sidebands'.

\begin{figure}[ht]
  \begin{center}
    \includegraphics[width=8.0cm,clip=on]{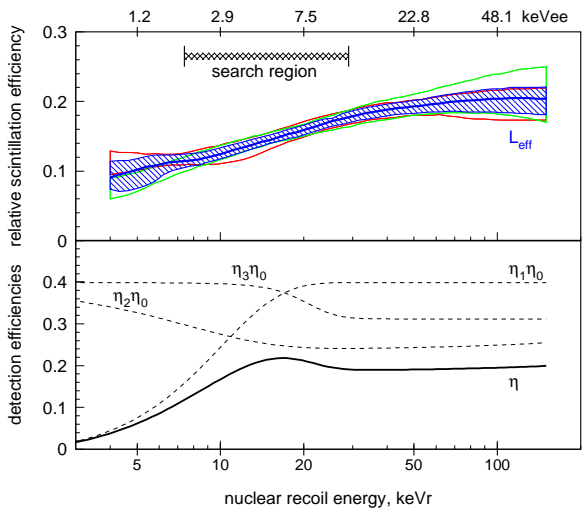}
  \end{center}
  \caption{\label{efficiency} Nuclear recoil efficiencies. Above:
relative scintillation efficiency for nuclear recoils in LXe at zero
electric field; the thick blue curve (adopted) combines the
measurements and uncertainties reported in Ref.~\cite{horn11} for FSR
data (68\% confidence region, in red) and SSR data (green). Below:
recoil detection efficiencies; $\eta$ subsumes a flat component
$\eta_0$=39.8\% (dominated by signal acceptance in S2/S1) and three
energy-dependent factors: $\eta_1$ is the S1 detection efficiency from
the 3-fold PMT coincidence required in software; $\eta_2$ is incurred
from quality cuts; $\eta_3$ relates to the timing cut on S1 pulses.}
\end{figure}

New measurements of the relative scintillation efficiency for nuclear
recoils in LXe ($L_{e\!f\!f}$) were derived from FSR and SSR
calibration data~\cite{horn11}. The decreasing $L_{e\!f\!f}$ curves
obtained therein were mutually consistent and agree, within
experimental errors, with recent neutron beam
data~\cite{manzur10,plante11}. For the purpose of converting between
electron and nuclear recoil energies we combined the two ZEPLIN-III
measurements into the curve shown in Figure~\ref{efficiency} (upper
panel). The SSR WIMP search region corresponds to 7.4--29~keVr.

The nuclear recoil detection efficiency comprises several
components. A set of constant factors combine to $\eta_0$=39.8\%,
including: DAQ livetime fraction (99.2\%), waveform quality cuts
(98.3\%), cuts on robustness of pulse parametrisation (96.9\%), veto
random coincidences (99.6\% and 99.0\% for the prompt and delayed
windows, respectively) and the above signal acceptance fraction
(42.7\%). Three additional curves, shown in Figure~\ref{efficiency}
(lower panel), describe the detection efficiency for S1 pulses (3-fold
requirement in software), the efficiency of the data quality cuts and
the additional timing cut on S1 signals for pulse shape
discrimination. The effective exposure for a 50~GeV/c$^2$ WIMP was
251~kg$\cdot$days.

\subsection{Experiment backgrounds}

Nuclear recoils from neutron single elastic scattering pose an
irreducible background to direct WIMP searches. In ZEPLIN-III,
detected radioactivity neutrons are most likely to arise in the
ceramic feedthroughs, the PMTs and the laboratory rock; muon-induced
neutrons contribute negligibly. The overall predicted rate is
3.05$\pm$0.5 events/year in 5--50~keVr assuming unity detection
efficiency, as detailed in Ref.~\cite{araujo11}. Translated to the SSR
effective exposure and search region this gives only
0.06$\pm$0.01~events in anti-coincidence with the veto.

A low-energy electron recoil background rate of 0.75~evts/kg/day/keVee
(`dru') was measured in the fiducial volume (with no quality cuts
applied), which represents a 20-fold reduction brought about by the
new phototubes. Monte Carlo predictions from a comprehensive inventory
of background sources, based on component-level radio-assays,
indicated 0.86$\pm$0.05~dru~\cite{araujo11}. Background electron
recoils are promptly tagged in the external veto with 28\% efficiency
below $\sim$100~keVee. This confirms that internal $\beta^-$ emitters
are insignificant, since the veto efficiency for coincident
$\gamma$-rays is practically identical to this value. Independently,
we measured the $^{85}$Kr decay rate at 7$\pm$2~mdru~\cite{araujo11}.

Assuming the discrimination factor achieved in the first run,
electron-recoil event leakage into the WIMP acceptance region should
represent $<$1~event, in line with the aspiration of a background-free
second run which motivated the experiment upgrades. However, analysis
of the prompt-vetoed $\gamma$-rays revealed that this was unlikely to
be the case. Background predictions (discussed below) based on
extrapolation of the electron recoil population into the signal box
and from low-energy $^{137}$Cs data indicated that a handful of
unvetoed events (7--9) were expected, confirming a loss of
discrimination power.

\section{WIMP-search results}
\label{results}

\begin{figure}[ht]
  \begin{center}
    \includegraphics[width=8.0cm,clip=on]{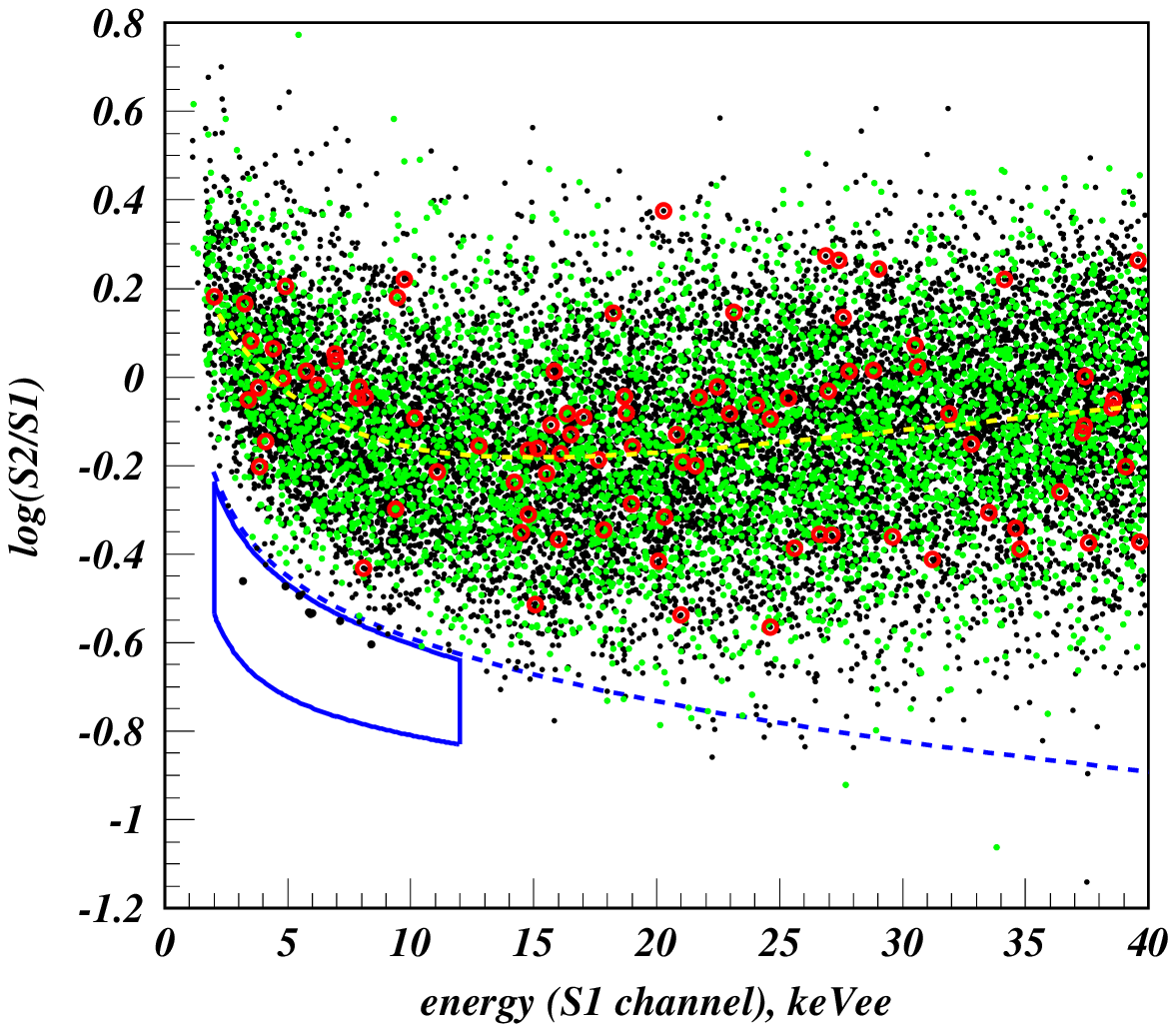}\\
    \includegraphics[width=8.0cm,clip=on]{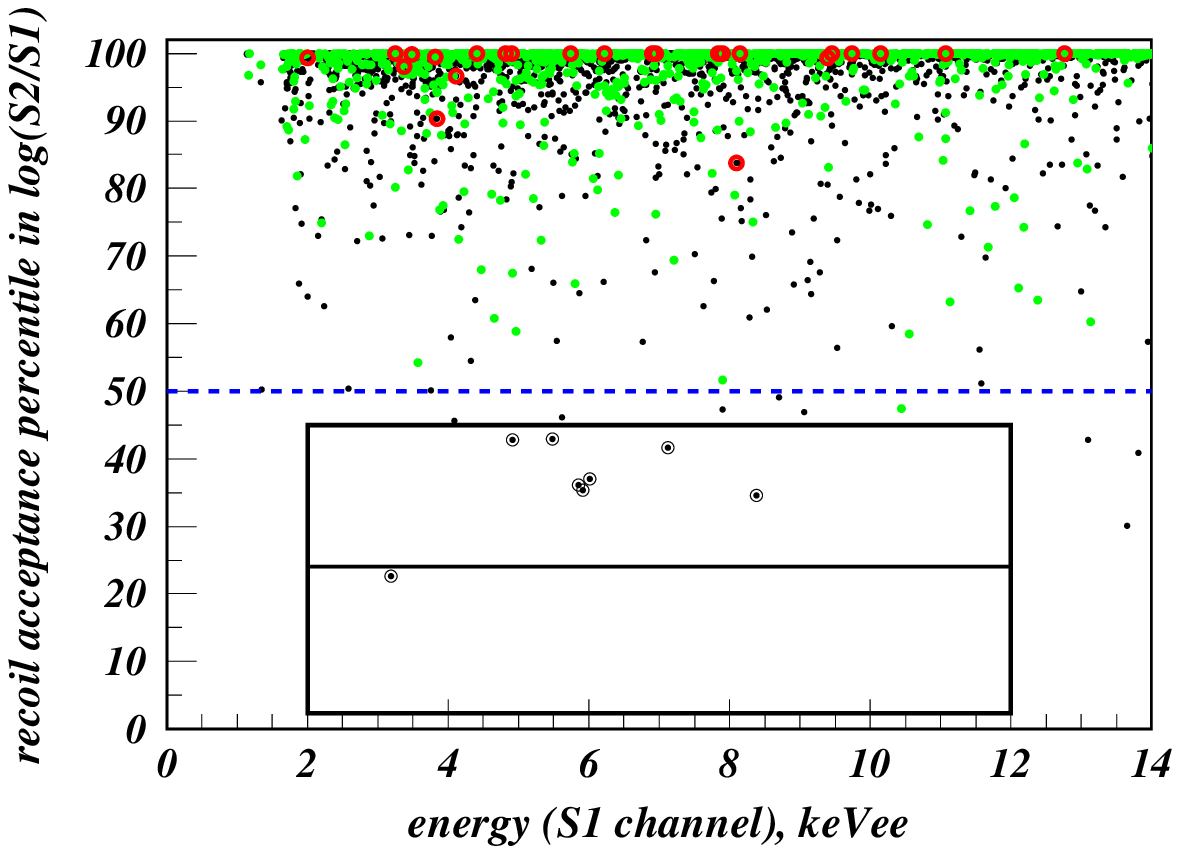}
  \end{center}
  \caption{\label{scatter} Above: fiducial events in full SSR
    exposure. Green markers label prompt veto coincidences (mostly
    $\gamma$-rays); events in the delayed window (mostly accidental
    coincidences) are in red. There are 8 unvetoed events in the WIMP
    acceptance region. The dashed lines show the nuclear and electron
    recoil band medians (in blue and yellow, respectively). Below:
    Distribution in signal acceptance (recoils from WIMPs and neutrons
    are distributed uniformly in the $y$-axis). There were seven (one)
    events in the upper (lower) region.}
\end{figure}

Upon unblinding, 12 events were observed in the acceptance region. A
detailed waveform inspection revealed larger than expected cross-talk
artifacts due to the poor electrical performance of the SSR PMTs:
gains differed by as much as 100 times within the array, and the extra
amplification required in some channels exposed contamination from the
higher gain channels (the array is powered by a single HV supply using
common dynode electrodes internally). Tighter cuts on the
goodness-of-fit of the reconstructed vertex were required to deal with
this issue (Figure~\ref{efficiency} already reflects this). The result
of the final (non-blind) analysis is shown in
Figure~\ref{scatter}. Eight events remained in the box. The event
reconstructed at 3.2~keVee in S1 and 1.1~keVee in S2 appears far below
the mean $\log_{10}$(S2/S1) for typical electron recoils with that S1
signal, but this is not necessarily anomalous: a median S2 signal of
1.1~keVee corresponds to electron recoils with only 0.6~keVee in S1
(cf.~yellow line in Figure~\ref{scatter}). Given 1.3~phe/keVee
scintillation yield, 0.8\% of all events with that S1 expected would
generate $\ge$4~phe due to Poisson fluctuation, thereby producing the
observed $\log_{10}$(S2/S1) ratio or lower. Considering the detected
rate of electron recoils, one such low-lying event is consistent with
background.

\begin{table*}[ht]
\begin{center}
\caption{Observations ($n_{obs}$), background estimates ($\mu_{b1,2}$)
and limits on the signal expectation ($\mu_s$) for the first (FSR) and
second (SSR) runs of ZEPLIN-III. Fiducial and net effective exposures
(50~GeV/c$^2$ WIMP) are presented along with relevant signal
acceptance parameters. Electron recoil background expectations are
estimated from Skew-Gaussian (SG) fits to data above the search region
in 2~keVee bins and also from $^{137}$Cs calibration. The 90\% CL
limit on the number of signal events in each run is derived with the
Profile Likelihood Ratio method.}
    \begin{tabular}{cccc|c|c|cc|c}
    \hline \hline
    run & kg$\cdot$days & \multicolumn{2}{c|}{acceptance} & $n_{obs}$ &
    neutrons & \multicolumn{2}{c|}{electron recoils, $\mu_{b2}$} & $\mu_s$ \\ 
    &  (net)      & keVr & fraction & & $\mu_{b1}$ & SG fit & Cs-137 & 90\% CL\\
    \hline
    FSR & 437.0   & 7--35 & 29--50\% & 4 & 0.5$\pm$0.3  & 5.2$\pm$3.1 & -- & $<$4.2\\
        & (107.3) &       &  2--29\% & 1 & 0.7$\pm$0.3  & 1.5$\pm$1.7 & -- &     \\
    \hline
    SSR & 1,343.8 & 7--29 & 24--45\% & 7 & 0.03$\pm$0.005& 5.5$\pm$2.2 & 8.3$\pm$2.9 & $<$5.1\\
        & (251.0) &       &  2--24\% & 1 & 0.03$\pm$0.005& 1.0$\pm$1.2 & 1.0$\pm$1.0 &     \\
    \hline \hline
    \end{tabular}
  \end{center}
\end{table*}

No {\em delayed} coincidences were recorded below the nuclear recoil
median and those registered above it are statistically consistent with
random coincidences. This allows us to set an upper limit of 0.75
neutron events for the search region (90\% CL), confirming the
successful mitigation of the neutron background. Two methods were used
to predict counts from electron recoil backgrounds; both are in good
agreement with the observation, suggesting no significant
signal. Binned Skew-Gaussian (SG) fits to the $\log_{10}$(S2/S1)
parameter above the search region, as in the FSR
analysis~\cite{lebedenko09a}, predict a total of 6.5$\pm$3.4 events
(cf. 8 observed). $^{137}$Cs calibration data provided independent
confirmation of this background within the available accuracy, with no
contamination from signal and no assumption of a functional dependence
for electron recoils; this yielded 9 events for 96\% equivalent
exposure.

To account for non-uniform background in $\log_{10}$(S2/S1) we
partitioned the search region in two; this was done explicitly to
maximise sensitivity given the SG background distributions. Note that
the contents of the search region were {\em not} considered in this
optimisation. To achieve this the partition was incremented
systematically and the background distribution was sampled by Monte
Carlo under the null hypothesis; the best sensitivity, calculated by
the Feldman-Cousins (FC) method~\cite{feldman98}, is reached at 24\%
acceptance. The observations in the resulting bins were seven and one
events, as shown in Figure~\ref{scatter} (lower). Table~1 confronts
observations and background estimates for the partitioned search
region.

Reanalysis of FSR data with updated SSR software, motivated by the new
$L_{e\!f\!f}$ measurements of Ref.~\cite{horn11}, revealed five events
in the FSR search region (which we kept unchanged from the original
analysis). Background expectations from SG fitting to these data are
also presented in Table~1. In this instance the $^{137}$Cs calibration
over-predicted the observations very significantly, which may be
caused by rate-dependent photocathode charging effects in this (FSR)
run~\cite{araujo04} or other systematic differences between the two
calibrations. We consider, however, that the SG function provides an
appropriate description of our background as validated by the SSR
$^{137}$Cs calibration. Once the non-negligible neutron contribution
is included, the FSR prediction is 7.9$\pm$4.8 (cf. 5 observed). The
same data-blind optimisation procedure led to a partition at 29\%; in
this run one event was just below that division in $\log_{10}$(S2/S1),
at 28.8\% acceptance.

\subsection{Signal inference}

A confidence interval for the signal expectation ($\mu_s$) in the SSR
acceptance region was obtained with a Profile Likelihood Ratio (PLR)
method which accounts for the uncertainty in the background
predictions (see, e.g.~\cite{sen09,rolke05}). We included estimators
of the nuisance parameter $\mu_{b2}$ from both the SG fits and the
$^{137}$Cs calibration; the latter is Poisson distributed and the SG
predictions are treated as Gaussian truncated at zero. The
distribution of profile likelihood ratio was determined by Monte Carlo
to ensure correct statistical coverage. The double-sided 90\% CL
interval for $\mu_s$ was $0-5.1$~events. The same procedure applied to
FSR data (without a $^{137}$Cs prediction in this instance) yielded
$0-4.2$~events with 90\% confidence.

These results are quite robust with respect to the particular choice
of binning and test statistic. For example, we conducted two-bin FC
calculations with background predictions capped at observation when
$\mu_b \!>\! n_{obs}$ (so as not to benefit from downward fluctuations
of background) and using a 10\% acceptance upper bin (as in the
original FSR analysis~\cite{lebedenko09a}). This yielded the same 4.2
events at 90\% CL for the first run and 4.8 for the SSR. These are
reassuringly close to the PLR results in Table~1, which treat
background uncertainties more formally.

\begin{figure}[ht]
  \begin{center}
  \includegraphics[width=8.5cm,clip=on]{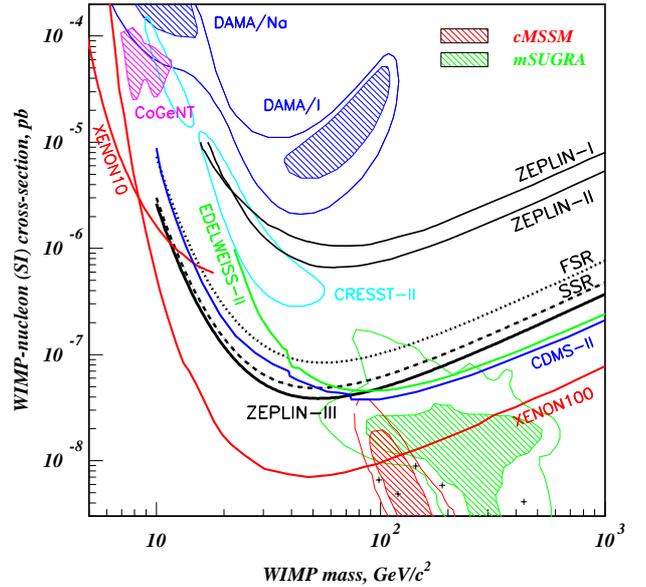}
  \end{center}
  \caption{\label{limit_si} 90\%-CL limits on WIMP-nucleon
  spin-independent cross sections from ZEPLIN-III (FSR, SSR and
  combined) and from XENON100~\cite{aprile11}, XENON10 (low energy
  analysis \cite{angle11}), CDMS-II~\cite{ahmed10b} and
  EDELWEISS-II~\cite{armengaud11}. Previous results from the ZEPLIN
  programme are also indicated~\cite{alner05,alner07a}.  In blue we
  represent the 3- and 5-$\sigma$ DAMA/LIBRA contours (2008 data, no
  ion channelling~\cite{barnabei08}) interpreted in
  Ref.~\cite{savage09}. The magenta contour is the fit to CoGeNT data
  under the light WIMP hypothesis~\cite{aalseth11}; in cyan is the
  2-$\sigma$ region from CRESST-II~\cite{angloher11}. The crosses are
  the original SUSY benchmark points~\cite{ellis02}. Favoured regions
  of parameter space from a 2008 Bayesian analysis in
  mSUGRA~\cite{trotta08} and the likelihood analysis of LHC data
  within cMSSM~\cite{buchmueller11a} are also shown.}
\end{figure}

Experimental upper limits on the scalar WIMP-nucleon elastic cross
section are shown in Figure~\ref{limit_si}, calculated with the
standard galactic halo model ($\rho_o$=0.3~GeV/c$^2$/cm$^3$,
$v_o$=220~km/s, $v_{e\!s\!c}$=544~km/s, and $v_{E}$=232~km/s)~and the
Helm form factor~\cite{helm56} parametrised in Ref.~\cite{lewin96}. A
minimum cross-section limit of 8.4$\times$10$^{-8}$~pb (90\% CL,
double sided) is reached at 55~GeV/c$^2$ WIMP mass for the FSR
(similar to the original result \cite{lebedenko09a} and slightly lower
than the result reported in Ref.~\cite{horn11} from a single-sided
`maximum patch' test statistic~\cite{henderson08}). The minimum of the
SSR curve is 4.8$\times$10$^{-8}$~pb, reached at
51~GeV/c$^2$. Adopting $L_{e\!f\!f}$ curves at the $\pm$1$\sigma$
levels from Figure~\ref{efficiency} does not affect the result at
curve minimum significantly, but the value at $10$~GeV/$c^2$ mass
varies in the range (2.2--9.3)$\times$10$^{-6}$~pb.

The combined result for the ZEPLIN-III experiment with mean
$L_{e\!f\!f}$, obtained from a four-bin PLR calculation which returns
$\mu_s$$<$6.0 events for the aggregate exposure, is also shown in the
figure; the curve minimum is 3.9$\times$10$^{-8}$~pb at 52~GeV/c$^2$.

\begin{figure}[ht]
  \begin{center}
  \includegraphics[width=8.5cm,clip=on]{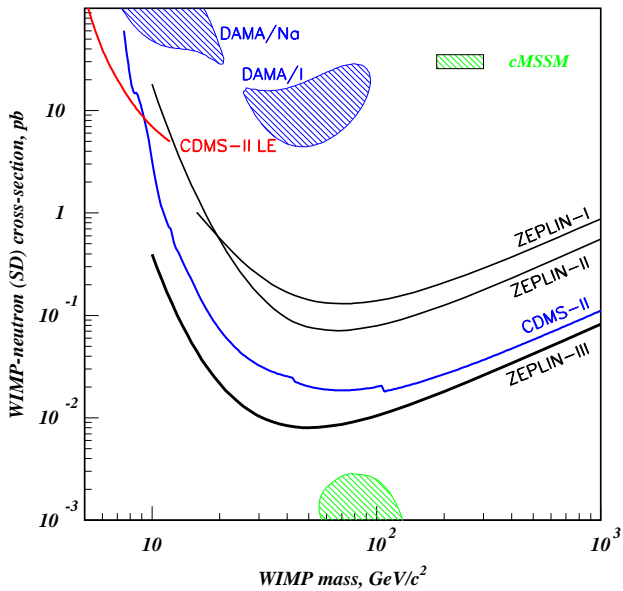}
  \end{center}
  \caption{\label{limit_sd} Limits on spin-dependent WIMP-neutron
   cross section (assuming no proton interaction) from ZEPLIN-III
   (FSR+SSR, with Bonn~CD potentials~\cite{toivanen08}) as well as
   CDMS-II (2004-09 data~\cite{ahmed09,ahmed10b} plus low-energy (LE)
   analysis~\cite{ahmed11}), ZEPLIN-I~\cite{alner05b} and
   ZEPLIN-II~\cite{alner07b}. Also shown is the 3-$\sigma$ DAMA
   evidence region (2008 data, no ion channelling~\cite{barnabei08})
   interpreted in Ref.~\cite{savage09}. The green hatched area is the
   tip of the 95\% probability region for cMSSM
   neutralinos~\cite{roszkowski07}.}
\end{figure}

Excellent sensitivity to spin-dependent WIMP-neutron interactions is
afforded by the odd-neutron isotopes $^{129}$Xe and $^{131}$Xe. The
spin-dependent result is calculated as described in
Ref.~\cite{lebedenko09b}, accounting for the composition of our xenon
(depleted in $^{136}$Xe) and using Bonn-CD nucleon-nucleon
potentials. The FSR+SSR combined curve, shown in
Figure~\ref{limit_sd}, has a minimum of 8.0$\times$10$^{-3}$~pb at
50~GeV/c$^2$ mass. At the time of writing there is no corresponding
result from XENON100; this is expected to be a few times lower. The
original XENON10 result \cite{angle08} is not shown in the figure
since this had assumed a constant scintillation efficiency and the
more favourable ($\sim$2) Bonn~A potential; a fair comparison would
raise this result to above the ZEPLIN-III curve.

\section{Conclusion}
\label{conclusions}

In this article we presented experimental upper limits on WIMP-nucleon
elastic scattering cross sections from the second run of ZEPLIN-III at
Boulby. These were derived from analysis of 1,344~kg$\cdot$days of
fiducial exposure acquired between June 2010 and May 2011. A 90\% CL
upper limit on the signal content from this exposure allows the
exclusion of a scalar WIMP scattering cross section above
4.8$\times$10$^{-8}$~pb/nucleon near 50~GeV/c$^2$ mass. The combined
result for the two runs constrains the scalar cross section to
3.9$\times$10$^{-8}$~pb and the WIMP-neutron spin-dependent cross
section to 8.0$\times$10$^{-3}$~pb. Along with
XENON100~\cite{aprile11}, XENON10~\cite{angle11},
CDMS-II~\cite{ahmed10b} and EDELWEISS-II~\cite{armengaud11} these
results disfavour an interpretation of DAMA in terms of nuclear
recoils from WIMPs as well as recent results from CRESST-II under the
canonical dark halo~\cite{angloher11}. Models favoured by CoGeNT
\cite{aalseth11} are harder to rule out completely from our data.

This second run followed the upgrade of the experiment with a new
array of purpose-developed, low-radioactivity PMTs, a veto detector
based on Gd-loaded polypropylene and plastic scintillator, and new
calibration hardware. The automation of the system enabled a 319-day
run with excellent stability and reduced manpower underground.

The new photomultipliers were a critical item to reach the design
sensitivity of 1$\times$10$^{-8}$~pb$\cdot$year. Although their
radiological background was excellent (35--50~mBq/PMT), their
performance compromised the experimental sensitivity by a factor of
$\sim$4 (from reduction in fiducial mass, of signal acceptance
fraction, of cut efficiency and from poorer
discrimination). Electron-recoil leakage to below the nuclear recoil
median had been 7,800:1 in the FSR compared to only 280:1 in the SSR;
no other significant changes were made to the internal hardware.

The addition of a new veto detector proved very valuable, confirming
negligible neutron background and providing a useful unbiased sample
of $\gamma$-ray background for open analysis. In our experience,
high-efficiency veto systems will be extremely important in future
experiments, from the point of view of background reduction,
diagnostics and enhancing their discovery power~\cite{ghag11}.

ZEPLIN-III concludes a successful series of three different LXe-based
experiments operated at Boulby since the mid 1990s, with progressively
stronger electric fields applied to their active targets. ZEPLIN-I
exploited pulse shape discrimination at zero field and produced
world-leading results~\cite{alner05,alner05b}. Delivering 1~kV/cm to
the liquid target, ZEPLIN-II was the first two-phase xenon WIMP
detector in the world~\cite{alner07a,alner07b}. ZEPLIN-III operated at
nearly 4~kV/cm in the first run and achieved the best discrimination
of any xenon detector, along with competitive WIMP results in
general. As several systems are now being designed and constructed
around the world with tonne-scale fiducial masses, the ZEPLIN
programme can claim to have pioneered some of the techniques that
helped two-phase xenon become a leading technology in the race to
discover WIMPs.

\section{Acknowledgements} 

The UK groups acknowledge support from the Science \& Technology
Facilities Council (STFC) for the ZEPLIN-III project and for operation
of the Boulby laboratory. LIP-Coimbra acknowledges financial support
from Funda\c{c}\~{a}o para a Ci\^encia e a Tecnologia (FCT) through
project grant CERN/FP/116374/2010 and postdoctoral grants
SFRH/BPD/27054/2006, /47320/2008, /63096/2009 and /73676/2010. ITEP
acknowledge support from the Russian Foundation of Basic Research
(grant 08-02-91851 KO\_a) and SC Rosatom (H.4e.45.90.11.1059 from
10.03.2011). We are also grateful for support provided jointly to ITEP
and Imperial from the UK Royal Society. ZEPLIN-III was hosted by
Cleveland Potash Ltd at the Boulby Mine and we thank CPL management
and staff for their long-standing support. We also express our
gratitude to the Boulby facility staff for their dedication. The
University of Edinburgh is a charitable body registered in Scotland
(SC005336).

\bibliographystyle{elsarticle-num}
\bibliography{HAraujo.bib}

\end{document}